# Highly selective chiral discrimination in high harmonic generation by dynamical symmetry breaking spectroscopy

*Ofer Neufeld and Oren Cohen*

Physics Department and Solid State Institute, Technion - Israel Institute of Technology, Haifa 32000, Israel.

Authors e-mails: ofern@tx.technion.ac.il, oren@technion.ac.il

We propose and numerically demonstrate a new very robust and highly selective method for femtosecond time-resolved chiral spectroscopy using high harmonic generation (HHG). The method is based on dynamical symmetry breaking from chiral media, and relies only on intense electric-dipole transitions, and not on the interplay of electric and magnetic dipoles. The symmetry breaking results in the emission of a strong chiral signal in the form of otherwise 'forbidden' harmonics (i.e., that are not emitted from achiral media). The intensity of these symmetry-forbidden harmonics is directly correlated to the media's enantiomeric excess, yielding chiral selectivity. On the contrary, the strength of the 'allowed' harmonics is chiral-independent, hence they can be used as a reference to provide chiral selectivity from a single measurement, unlike previous time-resolved schemes that require multiple measurements. We demonstrate numerically 96% discrimination level from microscopic gas-phase emission, outperforming by far previous time-resolved methods (the selectivity should be further enhanced when the HHG process is phase matched). We expect the new method to give rise to precise table-top characterization of chiral media in the gas-phase, and for highly sensitive time-resolved ultrafast probing of dynamical chiral processes.

Chirality is a fundamental property of asymmetric systems that is abundantly observed in nature. Its analysis and characterization is of tremendous importance in multiple scientific fields, including particle physics, astrophysics, chemistry, and biology. For example, amino acids are generally chiral, as well as DNA and other biologically active molecules [1,2], making molecular chiral spectroscopy a necessity for modern drug design [3,4]. Chiral spectroscopy is therefore paramount, and novel spectroscopic methods are required to enhance signal strength and resolution, as well as to probe novel systems with ultrafast chiral dynamics.

Chiroptical techniques that rely on light-matter interactions are often used to detect and characterize molecular chirality [5]. Chiroptical techniques are generally divided to two types: (I) those that require interactions with the magnetic components of the light field, and (II) those that rely solely on the interactions with the electric components of the light field, which are typically much stronger. The latter usually provide higher selectivity, while the former usually provide an easier experimental setup. Accordingly, methods that require magnetic interactions are historically more common, and include optical rotation, circular dichroism (CD) absorption spectroscopy, and many more [5,6]. Nevertheless, in the last two decades highly selective electric-dipole only based approaches have been successfully implemented in liquids (see [5,7,8] and refs. therein) and very recently also in the gas-phase with several innovative methods: photoelectron CD (PECD) [5,9,10], Coulomb explosion imaging [11,12], and microwave three-wave mixing spectroscopy [13,14]. Another challenge in chiral spectroscopy is probing ultrafast dynamical chiral processes, i.e., time-resolving chiral signals. Femtosecond resolution was obtained with vibrational CD spectroscopy in chiral liquids [15], and PECD [10,16–18] and photoexcitation CD (PXCD) [19,20] in the gas-phase. Still, PECD and PXCD are quite complicated to set-up experimentally, and require angularly resolving the photoelectron spectrum.

Recently, it was also discovered that high harmonic generation (HHG) with a helically polarized pump is chirality sensitive [21–23]. In chiral HHG (cHHG), a chiral medium is irradiated with an intense laser field, and the emission spectrum is measured. cHHG is a very promising technique, since it naturally leads to all-optical femtosecond time-resolved signals, and it is accessible in the gas-phase by simple table-top setups [21]. However, current geometries of cHHG exhibit relatively low selectivity: ~1% experimentally with somewhat elliptically polarized driver [21], and predicted 2.5% with the so-called bi-circular driver [23]. This relatively low selectivity means for example, that if the degree of chirality of the medium is 40% then the observed discrimination is ~1%, which may already be within the noise level and therefore not detectable. The main reasons for this relatively low selectivity are: i) the discrimination is based on interactions involving both the electric and the magnetic components of the light field, and (ii) the chiral signal is found by comparing small differences between two or more harmonic spectra that are measured separately, which increases noise. Clearly, a method for cHHG with much larger discrimination would open many opportunities, both for characterizing chiral gas media and for exploring ultrafast chirality.

Here, we predict and numerically demonstrate the first ultra-chirality-selective cHHG scheme that: i) relies solely on electric-dipole transitions, ii) the chiral signal is embedded in the intensity of otherwise 'forbidden' harmonics (i.e. they are not emitted from achiral media, which means that the achiral signal is zero), and iii) making possible chiral-selectivity from a single HHG spectrum (single-shot) by using chiral-independent harmonics for reference. The scheme is based on dynamical symmetry breaking in chiral media, utilizing the fact that chiral media inherently breaks certain symmetries (e.g., reflections and inversions). The pump field on the other hand is specifically chosen to uphold these symmetries, resulting in forbidden harmonic selection rules. This leads to the emission of otherwise 'forbidden' harmonics from chiral media, where the harmonic intensity is directly correlated to the enantiomeric excess (ee). We present here three realizations of the method, based on: (A) a static reflection symmetry, (B) a dynamical reflection symmetry, and (C) a dynamical inversion symmetry. The (C) scheme is the most promising, robust, and probably easiest to implement experimentally. Remarkably, it reaches a selectivity of up to 96%, outperforming by far previous cHHG methods [21–23], as well as other time-resolved methods [10,15,16,18,19].

**Dynamical symmetry breaking in chiral media**

We first briefly review dynamical symmetries (DSs) and selection rules in HHG [24], and how these can be broken by chiral media. We focus on the microscopic response of a molecule to an intense laser pulse, $\vec{E}(t)$. Since the pulse has a femtosecond duration, the Born-Oppenheimer approximation is employed, and in this paper we also employ the dipole approximation (DA) unless stated otherwise. The microscopic Hamiltonian of a single molecule interacting with a laser field is then given in atomic units and in the length gauge by:

$$H_\Omega(t) = -\frac{1}{2}\sum_j \vec{\nabla}_j^2 + \frac{1}{2}\sum_{i \neq j} \frac{1}{|\vec{r}_i - \vec{r}_j|} + \sum_j V_\Omega(\vec{r}_j) + \sum_j \vec{E}(t) \cdot \vec{r}_j \qquad (1)$$

where $H_\Omega(t)$ is the time-dependent multi-electron Hamiltonian, $\vec{r}_j$ is the coordinate of the $j$'th electron, $\vec{\nabla}^2_j$ is the Laplacian operator with respect to $\vec{r}_j$, $V_\Omega(\vec{r})$ is the molecular potential, and $\Omega$ represents the molecular orientation (as that of a rigid body). The Hamiltonian in Eq. (1) describes the interaction of an oriented molecule with a laser pulse, which due to the nonlinear laser-matter interaction (the right-most term in Eq. (1)) generates new harmonic frequencies. Quantum mechanically, the emitted harmonics are expressed by the second order time-derivative of the molecular-induced polarization, which can be calculated as the expectation value of the dipole operator:

$$\vec{P}_\Omega(t) = -\langle \Psi_\Omega(t) | \vec{r} | \Psi_\Omega(t) \rangle \qquad (2)$$

where $\Psi_\Omega(t)$ is the full multi-electron time-dependent wave function for the orientation $\Omega$, and the integration is performed over all electronic and spin coordinates. The emitted harmonic spectrum is extremely sensitive to the presence of symmetries in $H_\Omega$. For example, only odd harmonics are emitted if $H_\Omega$ is invariant under a half-wave rotational DS [25], and only circularly polarized harmonics are emitted if $H_\Omega$ is invariant under an $n$-fold rotational DS (for $n>2$) [26–28]. More generally, selection rules are derived as constraints that are a consequence of the invariance of $H_\Omega$ with respect to a unitary spatio-temporal transformation (see ref. [24] which presents a general derivation through group theory for DSs).

When the media is non-oriented, the laser pulse interacts with all possible rigid-body orientations of the molecule uniformly. Therefore, the induced polarization from all orientations should be summed:

$$\vec{P}_{tot}(t) = \int \vec{P}_\Omega(t) d\Omega \qquad (3)$$

The interaction is described by an effective Hamiltonian for the orientation-averaged ensemble, $H(t)$, which exhibits a higher symmetry than $H_\Omega(t)$. As a consequence, many of the molecular properties are 'washed-out' in HHG experiments – the orientation averaged Hamiltonian is very often spherically symmetric. For instance, HHG from molecular $SF_4$ gas or atomic Argon gas results in the same selection rules [29]. This is also true for different valence orbitals, where the same selection rules are observed from atomic Helium with valence s-states, and atomic Ne with valence p-states (though relative peak intensities are different) [30–32]. It is important to note that if the orientation-averaged molecular potential is spherically symmetric, then selection rules arise only as a consequence of the DSs of the driving laser and not of the microscopic medium.

We shall now describe the rational of our proposal. Consider two types of ensembles: a spherically symmetric (achiral) ensemble, and a chiral ensemble. Contrary to the spherically symmetric ensemble that is invariant under any rotation, reflection, and inversion (i.e., under $O(3)$), the chiral ensemble is only invariant under rotations (i.e., $SO(3)$), and cannot sustain reflection or inversion symmetries. This is the definition of chiral media – it is chiral if and only if it is distinguishable from its mirror image. Hence, the chiral ensemble is homogenous, but non-isotropic, which we can use in order to distinguish between the ensembles. To achieve this separation, we suggest using pump fields that: (i) generate bright high harmonics, (ii) exhibit a reflection or inversion related DS that leads to forbidden harmonic selection rules in achiral media, which are easy to observe and are broken in chiral media, leading to 'forbidden' emission that is correlated to the ee. (iii) Do *not* exhibit other DSs that also lead to the same selection rule but are not broken by chiral media (such as rotational DSs). To elucidate this last requirement, consider the following counter example: within the DA, any monochromatic field (linearly, elliptically, or circularly polarized) exhibits a dynamical inversion symmetry that forbids even harmonic generation [24]. Thus, one may expect to measure even harmonics from chiral media driven by such a field. However, monochromatic fields also uphold trivial $180^0$ rotational DSs around their propagation axis, which are not broken by chiral media, and lead to the exact same selection rule. Consequently, HHG emission from a monochromatic pump cannot detect chirality within the DA. In this paper we simulate and present three geometries that uphold these requirements, and involve the following DSs:

$$\hat{\sigma}_{xy} \tag{4}$$

$$\hat{Z} = \hat{\sigma}_{yz} \cdot \hat{\tau}_2 \tag{5}$$

$$\hat{F} = \hat{\imath} \cdot \hat{\tau}_2 \tag{6}$$

We follow the notation in ref. [24], where $\hat{\sigma}_{ij}$ represents reflection about the $ij$ plane, $\hat{\tau}_2$ represents temporal translation by half of the fundamental period ($T$), and $\hat{\imath}$ represents spatial inversion. Hence, Eq. (4) describes a 'static' space-only reflection transformation about the $xy$ plane ($z \to -z$), $\hat{Z}$ in Eq. (5) is a dynamical reflection symmetry about the $yz$ plane ($x \to -x, t \to t - T/2$), and $\hat{F}$ in Eq. (6) is a dynamical inversion symmetry ($\vec{r} \to -\vec{r}, t \to t - T/2$). The three DSs in Eqs. (4)-(6) lead to different selection rules on the emitted harmonic spectrum from the spherically symmetric ensemble: $\hat{\sigma}_{xy}$ results in forbidden induced $z$-polarization components (longitudinal components [33]), $\hat{Z}$ results in forbidden odd harmonics with $y$-polarization, and $\hat{F}$ results in forbidden even harmonics [24]. These symmetries are all broken by chiral media, which will cause an emission of a chiral signal in the form of new harmonics.

**Chirality selective cHHG spectroscopy**

In this section we present three schemes for chiral discrimination based on cHHG symmetry breaking spectroscopy using the symmetries in Eqs. (4)-(6). We numerically explore these schemes by solving the time dependent Schrödinger equation (TDSE) for a model chiral superatom within the single active electron (SAE) approximation. The numerical details for the chiral potential, its eigenstates, and the integration of the TDSE are found the appendix A.1, A.2. In order to implement an orientation averaged chiral ensemble, we perform consecutive calculations spanning different orientations $\Omega$ for the chiral atom. For the achiral ensemble, identical calculations are also performed for the partner enantiomer. The orientations are chosen such that for each orientation the inverted and reflected orientations are also tested (for details see appendix A.3).

**A. 'Static' reflection symmetry breaking**

We start with the simplest case of the space-only reflection symmetry, $\hat{\sigma}_{xy}$ in Eq. (4). We consider the cross-linearly polarized $\omega$-$2\omega$ HHG scheme [34,35] with the following pump field:

$$\vec{E}_A(t;z) = A(t) \cdot E_0(e^{i(\omega t - kz + \phi)}\hat{x} + e^{i(2\omega t - 2kz)}\hat{y}) \tag{7}$$

where $\omega$ is the optical frequency, $k$ is the wave vector, $A(t)$ is a dimensionless flat-top envelope function, $E_0$ is the field amplitude, $\phi$ is an arbitrary phase, and the spatial dependence of the field in Eq. (7) is neglected in the DA (plane wave treatment). We note that the microscopic pump field slightly differs in different $z$-planes due to the altered phase factor, and also in areas far from the beam center that may experience different intensity ratios between the $\omega$ and $2\omega$ fields. This does not hamper the scheme, since $\hat{\sigma}_{xy}$ symmetry holds for any phase or intensity relation between the $\omega$ and $2\omega$ fields. This configuration is schematically illustrated in Fig. 1(a). Fig. 1(b) shows the microscopic $z$-induced polarization components from the chiral ensemble, where $z$-components survive orientation averaging. In contrast, the $z$-polarized spectrum from the achiral ensemble is zero, i.e., $z$-polarization components do not survive orientation averaging. This occurs within the DA, and due to a different accumulated phase by the partner enantiomers – an (R) enantiomer driven by an orientation $\Omega$ is equivalent to an (S) enantiomer driven by the reflected orientation $\Omega^*$, because the pump is reflection invariant. However, in the reflected orientation the $z$-components acquire an overall minus sign. Hence, in the achiral ensemble these contributions **exactly** interfere destructively (this is schematically illustrated in Fig. 2). On the other hand, there is no such effect in the chiral ensemble, because by definition there is no orientation for which the molecule can be superposed onto its mirror image. This also means that the cHHG spectra from partner chiral ensembles are indistinguishable except for a global phase (as should be within the DA), though the helicity can be detected by reconstructing the phase, or by simultaneously measuring the optical rotation. In addition, we note no interference between different HHG channels is required, because the mechanism responsible for discrimination is a result of the coherent interference between emission from different molecular orientations. Fig. 1(c) further presents the integrated intensity of $z$-induced polarization components per harmonic order vs. the medium's degree of chirality (DOC), which is numerically calculated by coherently adding the polarization from both enantiomers and renormalizing:

$$\vec{P}_{tot}^{\chi}(t) = \frac{\left(\vec{P}_{tot}^{(R)}(t) + (1-\chi)\vec{P}_{tot}^{(S)}(t)\right)}{2-\chi} \tag{8}$$

where we define $\chi$ as the DOC ranging from 1 for the chiral ensemble to 0 for the achiral ensemble, and $\chi$ is related to the ee through: $ee = \chi/(2-\chi)$. The intensity of the longitudal induced polarization increases exponentially as a function of $\chi$, which results in extremely high selectivity that is effectively single-shot (because the intensity of the 'forbidden' emission can always be normalized with respect to the allowed emission that is insensitive to the chirality). We define the discrimination between the two ensembles per harmonic order $n$:

$$S_n = \frac{I_n^{\chi=1} - I_n^{\chi=0}}{I_n^{\chi=1} + I_n^{\chi=0}} \tag{9}$$

, which ranges from 0 to 100% (the definition in Eq. (9) is not the standard nomenclature in the field [5,21–23] where $S_n$ is often multiplied by a factor 2 hence it is not normalized to 100%, making it less suitable in our case). Other effective observables for discrimination can also be defined. For example, one may average $S_n$ over all harmonic orders, which of course increases the signal to noise ratio. Numerically, we find exactly 100% discrimination between the chiral and achiral media for the longitudal spectrum. This is because the signal to noise ratio in our numerical implementation is as high as machine precision, which is attributed to the fact that $\hat{\sigma}_{xy}$ is an exact symmetry of our model (this symmetry has no temporal part and is therefore not broken by the finite duration of the laser pulse, or by ionization). This is also a verification that the ensemble is indeed orientation averaged. In reality, magnetic interactions that are neglected in this analysis may increase the noise in accordance with standard cHHG (~2.5%) [23], possibly slightly reducing the selectivity.

Surprisingly, the pump field in Eq. (7) leads to chiral selectivity even though it is comprised from linearly polarized fields. This may seem like a contradiction, since chirality may only be detected through an interaction with another chiral object. However, the cross-linearly polarized field is indeed helical and has a non-zero optical chirality [36–38], breaking time-reversal symmetry.

Notably, in this scheme the longitudal modes that carry the chiral signal are non-propagating, and therefore are challenging to detect. We next present two other schemes where the chiral signal is embedded in propagating modes.

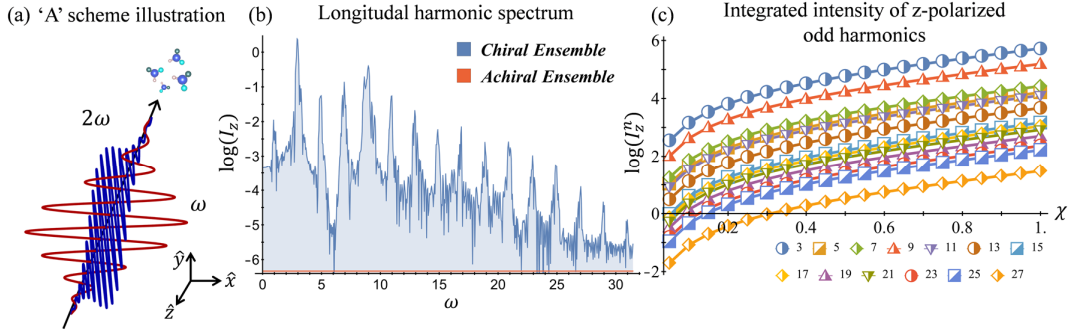

Fig. 1 – 'Static' reflection symmetry breaking based chiral selectivity in HHG: numerical results. (a) Illustration of scheme 'A' with the pump field $\vec{E}_A(t)$ from Eq. (7). (b) Numerically calculated $z$-induced (longitudinal) high harmonic emission from the chiral/achiral ensembles ($\psi_6$), for $\lambda$=900nm, $I_{max}$=6×10$^{13}$ [W/cm$^2$], $\phi = \pi/3$, and a trapezoidal envelope with 7 optical cycle long turn-on/off and 7 optical cycle long flat-top. (c) Integrated power per harmonic of the $z$-polarized emission vs. the medium's degree of chirality in log scale.

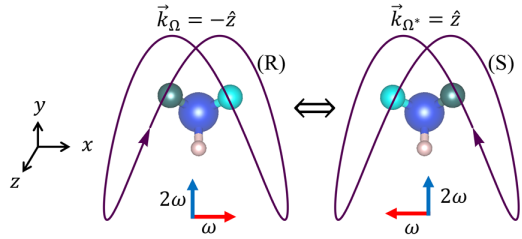

Fig. 2 – Mechanism for destructive interference in achiral HHG from scheme 'A': for every orientation $\Omega$ of enantiomer (R), a dipole approximation equivalent orientation $\Omega^*$ for enantiomer (S) exists due to the reflection symmetry of the pump. These two orientations exactly destructively interfere for longitudinal components in the achiral ensemble. However, by definition no two such configurations exist in the chiral ensemble. Red and blue arrows represent polarization directions of fundamental and second harmonic beams, the purple Lissajou plot shows the overall vector pump field in the dipole approximation, and $\vec{k}$ represents the propagation direction in each configuration.

## B. Dynamical reflection symmetry breaking

We now consider chiral symmetry breaking of the dynamical reflection symmetry in Eq. (5). In order to engineer a laser field that has $\hat{Z}$ symmetry and upholds the above requirements one must consider non-collinear HHG schemes [39–41]. Any collinear scheme necessarily exhibits dynamical 180$^0$ rotations that prevent chiral selectivity in the propagating modes. Therefore, we consider the following bi-chromatic $\omega$-2$\omega$ cross-beam geometry:

$$\vec{E}_B(t;x,z) = A(t) \cdot E_0(e^{i(\omega t+kz+\phi)}\hat{x} + \Delta \cdot e^{i(2\omega t-2kx)}(\hat{y} + i\varepsilon_2\hat{z})) \quad (10)$$

where $\Delta$ is the relative amplitude ratio between the $\omega$ and 2$\omega$ fields, $\varepsilon_2$ is the ellipticity of the 2$\omega$ beam, and other symbols are as previously defined. The field in Eq. (10) is comprised of a fundamental $\omega$ beam propagating in the $-z$ direction and linearly polarized along the $x$-axis, and a second 2$\omega$ beam propagating in the $x$ direction and elliptically polarized in the $yz$ plane. As before, $\hat{Z}$ DS is maintained for any phase relation $\phi$, ellipticity $\varepsilon_2$, and amplitude ratio $\Delta$, but we must uphold $\Delta, \varepsilon_2 \neq 0$, such that in the focus of both beams we don't have a local field with rotational DSs that lead to identical selection rules. This scheme is illustrated in Fig. 3(a).

Fig. 3(b) plots the intensity of the emitted harmonics polarized along the $y$-axis from the chiral and achiral ensembles, which conserve momentum for all of the harmonic orders and are thus propagating modes. As seen, $y$-polarized odd harmonics do not survive orientation averaging in the achiral ensemble due to the selection rules imposed by $\hat{Z}$ DS. However, they do survive orientation averaging in the chiral ensemble. The mechanism for this effect is identical to that in the static symmetry breaking case described in Fig. 2, though here the signal to noise ratio is lower since $\hat{Z}$ DS is also somewhat broken by the finite duration of the pulse and by ionization in the medium (i.e., some odd harmonics can also be emitted from the achiral ensemble). Note that in Fig. 3(b), the intensity of the even 'allowed' harmonics does not depend on the chirality, and can be used as a normalization to make the scheme single-shot. Fig. 3(c) presents the integrated power of $y$-polarized odd harmonics that depends near exponentially on $\chi$, making this scheme highly selective. We find a discrimination of 99.3% between the chiral and achiral ensembles through the integrated $y$-polarized odd harmonic power. We expect the discrimination can be enhanced even further using phase matching, though magnetic interactions will slightly decrease this selectivity (by ~2.5% according to standard cHHG [23]).

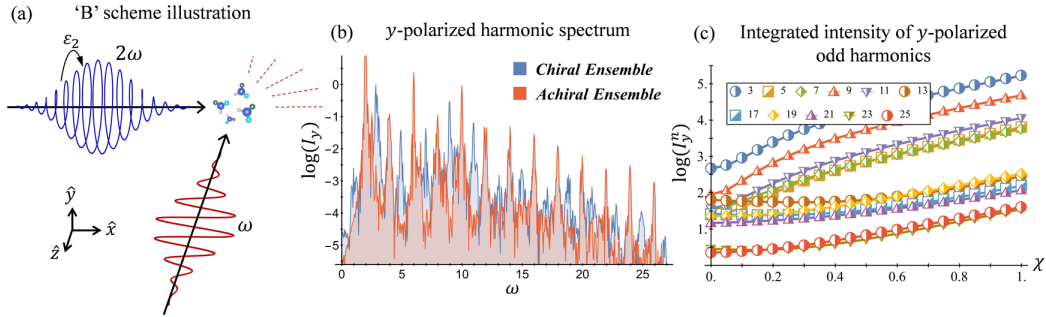

Fig. 3 – Dynamical reflection symmetry breaking based chiral selectivity in HHG: numerical results. (a) Illustration of scheme 'B' with the pump field $\vec{E}_B(t)$ from Eq. (10). (b) $y$-polarized high harmonics emitted from the chiral/achiral ensembles ($\psi_6$), for fundamental wavelength $\lambda = 900nm$, $I_{max}=4.5\times10^{13}$ [W/cm$^2$], $\phi = \pi/3$, $\Delta = 1$, $\varepsilon_2 = 0.25$, and a trapezoidal envelope with 7 cycle turn-on/off and 7 cycle flat-top. (c) Integrated power of the $y$-polarized odd harmonics vs. the medium's DOC in log scale.

## C. Dynamical inversion symmetry breaking

We consider a third method for chiral spectroscopy through dynamical inversion symmetry breaking (Eq. (6)) that is driven by the following non-collinear bi-chromatic $\omega$-$3\omega$ geometry:

$$\vec{E}_C(t;\vec{r}) = A(t) \cdot E_0 \left( e^{i(\omega t - \vec{k}_1 \cdot \vec{r} + \phi)}(\hat{e}_1 - i\varepsilon_1 \hat{y}) + \Delta \cdot e^{i(3\omega t - \vec{k}_3 \cdot \vec{r})}(\hat{e}_3 + i\varepsilon_3 \hat{y}) \right) \quad (11)$$

where $\vec{k}_1 = k(\sin(\alpha)\hat{x} - \cos(\alpha)\hat{z})$, $\vec{k}_3 = -3k(\sin(\alpha)\hat{x} + \cos(\alpha)\hat{z})$, $\hat{e}_1 = \cos(\alpha)\hat{x} + \sin(\alpha)\hat{z}$, $\hat{e}_3 = \cos(\alpha)\hat{x} - \sin(\alpha)\hat{z}$, $k$ is the magnitude of the wave vector related to the optical frequency, $\varepsilon_1$ is the ellipticity of the $\omega$ beam, $\varepsilon_3$ is the ellipticity of the $3\omega$ beam, $\alpha$ is the opening angle between the two beams within the $xz$ plane, and other symbols are as previously defined (see Fig. 4(a) for illustration). As in the previous cases, $\hat{F}$ symmetry is maintained for all coordinates along the beam paths, for any ellipticities $\varepsilon_{1,3}$, amplitude ratio $\Delta$, opening angle $\alpha$, phase $\phi$, as well as in the presence of $\omega$ components in the $3\omega$ beam. In order to prevent other DSs that lead to the same selection rules we must uphold: $\alpha, \Delta, \varepsilon_{1,3}, \phi \neq 0$.

Fig. 4 (b) shows that $\hat{F}$ symmetry is broken by the chiral ensemble and leads to the generation of even harmonics, while this symmetry is upheld in the achiral ensemble where even harmonics are forbidden. Fig. 4(c) shows the peak intensity of emitted even harmonics as a function of the medium's DOC, which similarly to previous sections behaves near exponentially. A discrimination of 96.6% is obtained by summing the even harmonic peak intensities. Notably, there are 7 independent degrees of freedom in the bi-chromatic beam $\vec{E}_C(t)$, including ellipticities, opening angles, etc., which allow a large freedom for optimizing the signal to noise ratio. Due to limited computational time we did not optimize these parameters (each calculation for a single orientation takes ~24hrs on a Tesla K80 GPU, and many calculations need to be performed to orientation average both ensembles). We only note that the parameters should be chosen as to keep the pump field in Eq. (11) far from exhibiting rotational DSs that lead to the same selection rules. Notably, $\hat{F}$ symmetry may be obtained by any pair of beams with frequency ratios $\omega_1:\omega_2$ that in reduced fractional form are odd integers, i.e., $\frac{\omega_1}{\omega_2} = \frac{n_1}{n_2}$, where both $n_1$ and $n_2$ are odd.

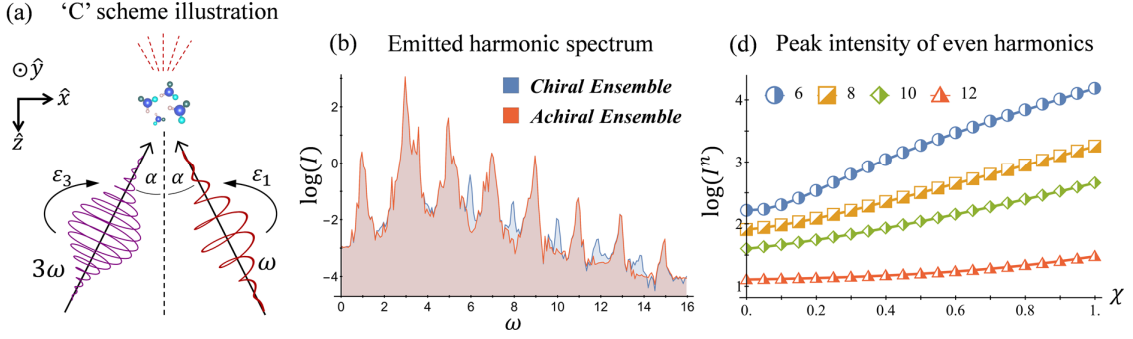

Fig. 4 – Dynamical inversion symmetry breaking based chiral selectivity in HHG: numerical results. (a) Illustration of scheme 'C' with the pump field $\vec{E}_C(t)$ given in Eq. (11). (b) High harmonics emitted from the chiral/achiral ensembles ($\psi_{15}$), fundamental wavelength $\lambda$=700nm, $I_{max}$=10$^{13}$ [W/cm$^2$], $\phi = \pi/2$, $\Delta = 1$, $\varepsilon_1 = 0.4$, $\varepsilon_2 = 0.3$, $\alpha = 20^0$, and a trapezoidal envelope with 4 cycle turn-on/off and 5 cycle flat-top. (c) Peak power of the even harmonics emitted from a chiral medium vs. the medium's DOC in log scale.

## Conclusions

In this paper, we proposed and numerically demonstrated a novel general approach for cHHG that is based on dynamical symmetry breaking spectroscopy (and not on the detailed mechanism of the HHG process). The scheme relies solely on intense electric-dipole transitions (not on the interplay of both magnetic and electric dipoles, as all other cHHG schemes [21–23]), making it highly robust. We explored three different realizations for the scheme by engineering pump fields with various dynamical symmetries. For example, the third approach we presented is based on dynamical inversion symmetry breaking, which has many optimizable degrees of freedom and is relatively simple to set-up. Remarkably, it results with extremely high chiral selectivity of up to 96%, outperforming by far other time-resolved techniques [10,15,16,18,19,21–23]. Phase matched propagation should further enhance the selectivity. Our work paves the way for table top highly sensitive chiral spectroscopy in the gas-phase, and for single-shot femtosecond time-resolved ultrafast spectroscopy of chiral processes. We also believe that this work will advance HHG spectroscopy for other physical phenomenon, including magnetic interactions, spin-orbit effects, and more, by implementing an analog approach that is based on DS breaking.

## Acknowledgements

This work was supported by the Israel Science Foundation (grant no. 1225/14), and the Wolfson foundation. O.N. gratefully acknowledges the support of the Adams Fellowship Program of the Israel Academy of Sciences and Humanities.

## Comment

While finalizing the manuscript, we learned through private communication with Olga Smirnova that her group is also developing cHHG that is based on only electric-dipole interaction.

## Appendix

### A.1 Chiral potential – numerical details:

A chiral superatom was described by the following model chiral core potential:

$$V(\vec{r}) = -\frac{z_1}{\sqrt{(\vec{r}-\vec{r}_1)^2 + a}} - \frac{z_2}{\sqrt{(\vec{r}-\vec{r}_2)^2 + a}} - \frac{z_3}{\sqrt{(\vec{r}-\vec{r}_3)^2 + a}} - \frac{z_4}{\sqrt{(\vec{r}-\vec{r}_4)^2 + a}} \qquad (12)$$

where $z_{1,2,3,4} = 1.5, 1.25, 1, 0.75$, respectively, $\vec{r}_1 = 0$, $\vec{r}_2 = -\hat{x}$, $\vec{r}_3 = \hat{y}$, $\vec{r}_4 = \hat{z}$ bohr, and $a = 0.05$ a.u. The iso-surface plot for this potential is presented in Fig. 5(a). The enantiomeric atom is found by reflecting along the $yz$ plane (i.e., $\vec{r}^*_2 = \hat{x}$). The eigenstates for this potential are found by complex time propagation of the TDSE (as detailed in A.2) and a Gram-Schmidt algorithm, implemented to a self-consistency convergence level of 10$^{-8}$ hartree, and 10$^{-5}$ in the maximal wave function difference. The initial states chosen in calculations are the 6$^{th}$, 14$^{th}$, and 15$^{th}$ excited states. Their ionization potentials are: Ip= 1.112, 0.776, 0.633 hartree, respectively, which were converged in the grid parameters up to 10$^{-3}$ hartree. The iso-surface plots for these orbitals are presented in Fig. 5(b, c, d), and clearly show a chiral nature (the lack of any reflection or inversion symmetries). Notably, the main script above presents numerical results for this model chiral superatom even though its orbitals are not highly chiral (see Fig. 5 – the orbitals do not exhibit exact reflection or inversion symmetries, but are not far from it). In a full multi-electron molecular system with valance orbitals that are localized on the chiral center, we expect the discrimination to be even higher. This prediction is supported by numerical calculations from different initial orbitals that show orbitals with larger chirality lead to larger discrimination. This also numerically shows the importance of the chirality of the initial state, and not just that of the chiral molecular potential.

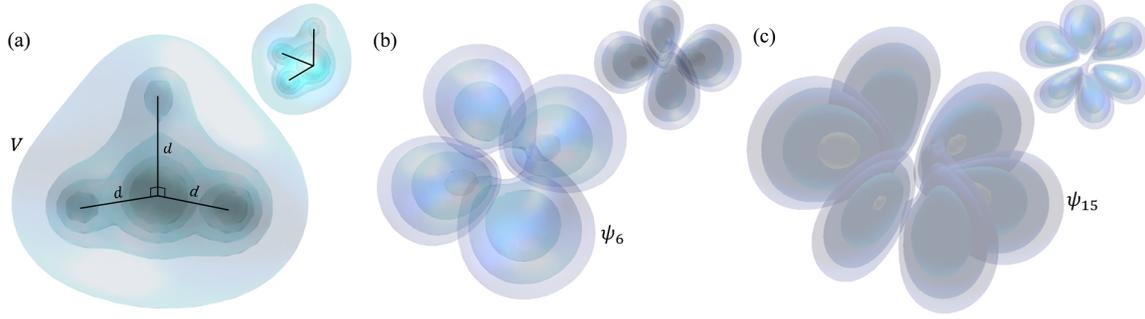

Fig. 5 – Iso-surface plots for: (a) the chiral atomic potential, (b) the 6$^{th}$ orbital, and (c) the 15$^{th}$ orbital. In each sub-figure a second viewpoint is shown in inset. Iso-surfaces for the potential are: $s = 0.15, 0.3, 0.4, 0.5, 0.6 V_{max}$, and for the orbitals are: $s = 0.05, 0.1, 0.25, 0.75 |\psi|^2_{max}$. In (a) lines represent the molecular backbone for the model potential, where $d = 1 a.u.$

## A.2 TDSE – numerical details:

The TDSE defined by the Hamiltonian in Eq. (1) was solved numerically on a 3D real space L$^3$ grid for L=100 bohr, with equidistant grid spacing $\Delta x = \Delta y = \Delta z = 0.2604$ bohr. The numerical integration was performed using the 3$^{rd}$ order split step method [42,43], with a time step of $\Delta t = 0.02$ a.u. if $\lambda = 900 nm$ and $\Delta t = 0.015$ a.u. if $\lambda = 700 nm$. An imaginary absorbing potential was used to prevent reflection from the boundaries of the form:

$$V_{ab}(\vec{r}) = -i\eta(|\vec{r}| - r_0)^\gamma \Theta(|\vec{r}| - r_0) \qquad (13)$$

where $\eta = 1.5 \times 10^{-3}$, $r_0 = 36$ bohr, and $\gamma = 4.3$ for $\psi_{15}$ and $\eta = 10^{-3}$, $r_0 = 33$ bohr, and $\gamma = 4$ for $\psi_6$. The induced polarization was calculated according to Eq. (2) on a twice reduced time grid (i.e. every other time step), and the dipole acceleration was calculated directly using a 5$^{th}$ order finite difference approximation for the second order time-derivative:

$$\vec{a}(t) = \frac{d^2}{dt^2} \vec{P}_{tot}(t) \qquad (14)$$

## A.3 Orientation averaging – numerical details

Orientation averaging was achieved by repeatedly solving the TDSE for varying beam propagation orientations, and realigning the calculated induced polarization along the correct axis with rotation matrices. This was done for 24 major alignments along the three Cartesian axes, which were pre-chosen such that for each orientation Ω in the ensemble, the inverted and reflected orientations also exist.

Two independent variables are required in order to define the orientation of a rigid body in 3D. We used a normalized vector $\vec{v}$, that defines the $z$-axis in the frame of the propagating beams in terms of the Cartesian coordinates $x', y', z'$ in the molecular frame of reference (which corresponds to the potential form in Eq. (12)), accompanied by an angle $0 < \theta \leq 2\pi$ that defines the relative rotation about the $\vec{v}$ axis (such that $\theta = 0$ corresponds to the $x$-axis in the beam's frame of reference). The 24 orientations are given by all permutations of the angles $\theta = 0^0, 90^0, 180^0, 270^0$, and $\vec{v}' = \pm \hat{x}', \pm \hat{y}', \pm \hat{z}'$, which all together span 24 rigid body orientations.